\documentclass{aa}
\usepackage{graphicx}
\def\ltsima{$\; \buildrel < \over \sim \;$}
\def\simlt{\lower.5ex\hbox{\ltsima}}
\def\gtsima{$\; \buildrel > \over \sim \;$}
\def\simgt{\lower.5ex\hbox{\gtsima}}
\def\cgs{{erg cm$^{-2}$ s$^{-1}$}}
\def\ergs{{erg s$^{-1}$}}
\def\cm2{{cm$^{-2}$}}
\def\xdof{{$\chi^{2}$(dof)}}
\def\xrd{{$\chi^{2}_{\rm \nu}$(dof)}}

\def\xnn{{$\chi^{2}_{\rm \nu}$}}
\def\fhx{{$F_{\rm 2-10}$}}

\def\lum{{$L_{2-10}$}}
\def\p1{{Paper I}}
\def\fsx{{$F_{\rm 0.5-2}$}}
\def\xmm{{\em XMM--Newton}}
\def\chandra{{\em Chandra}}

\def\chandra{{\em Chandra}}

\def\xmm{{\em XMM--Newton}}

\def\nh{{N$_{\rm H}$}}
\def\epic{{\em EPIC}}

\def\mos{{\em MOS}}
\def\pn{{\em PN}}

\def\f14{{10$^{-14}$}}
\def\f13{{10$^{-13}$}}
\def\f12{{10$^{-12}$}}
\def\f11{{10$^{-11}$}}

\def\4u{{4U~1344$-$60}}

\def\feka{{Fe K$\alpha$}}
\def\3c{{3C 234}}
\def\mgx{{Mg$^{7+}$/X}}
\begin{document}

\title{Heavy absorption and soft X-ray emission lines in the \xmm~spectrum
  of the Type 2 radio-loud quasar 3C 234}
\author{E.~Piconcelli\inst{1}, S.~Bianchi\inst{2}, G.~Miniutti\inst{3}, F.~Fiore\inst{1},
  M.~Guainazzi\inst{4}, E. Jimenez-Bailon\inst{5}, G.~Matt\inst{2}}
\titlerunning{XMM-Newton observation of the Type 2 quasar 3C 234}\authorrunning{E.~Piconcelli et al.}
\offprints{Enrico Piconcelli, \email{piconcelli@oa-roma.inaf.it}}
\institute{Osservatorio Astronomico di Roma (INAF), Via Frascati 33, I--00040
  Monteporzio Catone (Roma), Italy  \and Dipartimento di Fisica, Universit\`a degli Studi Roma 3, Via della Vasca Navale 84, I--00146 Roma, Italy \and Institute of Astronomy, Madingley Road, Cambridge CB3 0HA, UK \and European Space  Astronomy Center of ESA, Apartado 50727, E--28080 Madrid,
  Spain \and Instituto de Astronom\'ia, UNAM, Apartado 70264, 04510 Ciudad de M\'exico, Mexico}
\date{}

\abstract{}{We report results on a 40 ks \xmm~observation of the Type 2 quasar 3C 234.
Optical spectropolarimetric data have demonstrated the presence of a hidden broad-line region in this powerful (M$_V$
$\leq$ $-$24.2 after reddening and starlight correction) narrow-line FRII radio galaxy.
Our analysis is aimed at investigating the X-ray spectral properties of this peculiar source which have 
remained poorly known so far.}
{We analyze the 0.5--10 keV spectroscopic data collected by the \epic~cameras in 2006.}
{The X-ray spectrum of this radio-loud quasar is typical  of a local Compton-thin Seyfert 2 galaxy.
It exhibits strong absorption (\nh~$\sim$ 3.5 $\times$
10$^{23}$ \cm2) and  a narrow, neutral 
\feka~emission line with an equivalent width of $\approx$140$\pm$40 eV.
Our observation also reveals that the soft portion of the spectrum is 
characterized by strong
emission lines with a very low level of scattered primary continuum.
A possible explanation of these features in terms of thermal emission from a
two-temperature collisionally ionized plasma emission seems to
be unlikely due to the high luminosity estimated for this component ($L_{0.5-2}$ $\sim$ 6
$\times$ 10$^{42}$ \ergs).
It is likely that most
of the soft X-ray emission originates from a photoionized plasma
 as commonly observed in  obscured, radio-quiet Seyfert-like AGNs.}
 {This X-ray observation has definitively confirmed the presence of a hidden quasar in 3C 234.
The line-rich spectrum and the steepness of the hard X-ray continuum ($\Gamma$ $\approx$ 1.7) found in this  source  weaken the hypothesis that the bulk of
the X-ray emission in radio-loud AGNs with high excitation optical lines 
arises from jet non-thermal emission.}
\keywords{Galaxies:~active --
  Galaxies:~nuclei -- Quasars:~individual:~3C 234 -- X-ray:~galaxies}
   \maketitle

\section{Introduction}

Type 2 quasars (QSO2s hereafter) are usually defined as those active galactic nuclei
with the following properties: (i) a highly (i.e. \nh$\geq$10$^{22}$ \cm2) absorbed
X-ray emission (ii) an intrinsic, hard X-ray (i.e. 2-10 keV) luminosity $\geq$
10$^{44}$ \ergs, and (iii) the lack of broad emission lines in their optical/UV
spectra (the latter condition is often overcome since these objects are
usually too weak, i.e. $R$\simgt~25 mag, for  optical spectroscopy).
QSO2 are therefore the luminous counterpart of Seyfert 2 galaxies.
Even if their existence was postulated by the AGN unification model (Antonucci
1993; Urry \& Padovani 1995) QSO2 have represented for many years the
long-sought missing AGN population (Halpern et al. 1999). 

Optical and soft X-ray (0.5-2 keV) surveys, being strongly biased by
obscuration, fail to detect clearcut examples of such objects.   However,
over the past $\sim$10 years, a very limited number of genuine QSO2s have been
discovered (e.g. Brandt et al. 1997; Franceschini et al. 2000; Norman et
al. 2002; Stern et al. 2002; Della Ceca et al. 2003; Iwasawa et al. 2005;
Gandhi et al. 2006) thanks to X-ray observations above 2 keV.  In particular,
the recent deep and wide-area hard X-ray surveys performed with  \chandra~and
\xmm~(e.g. Mainieri et al. 2002; Fiore et al. 2003; Brandt \& Hasinger 2005, Eckart et al. 2006; Wang
et al. 2007; Georgantopolous et al. 2007; Lacy et al. 2007)  have efficiently
detected several dozens of QSO2 candidates.  Unfortunately, most of them are
sources with weak optical emission, showing a very high
X-ray--to--optical flux ratio (X/O $>$ 10), and, therefore,  with no reliable
information about their redshift and classification (e.g. Maiolino et
al. 2006).  Furthermore, these hard X-ray surveys 
have established that the obscured AGN fraction decreases with increasing hard
X-ray luminosity (Ueda et al. 2003; Steffen et al. 2003; La Franca et
al. 2005; Akylas et al. 2006)  and strongly increases with the hard X-ray
flux only at \fhx~\simlt10$^{-14}$ \cgs~(Piconcelli et al. 2003).  The combination
of these findings explains the difficulties met so far  in collecting a large
sample of intrinsically-luminous, heavily obscured AGNs.  According to the
most up-to-date version of the synthesis model of the cosmic X-ray background
(CXB), e.g. Gilli, Comastri \& Hasinger (2007),  the population of obscured
QSOs accounts for $\sim$15\% of the CXB in the 2--10 keV band.

Selecting sources with large [OIII]
luminosity (i.e. L$_{\rm [OIII]}$ \simgt~10$^{42}$ \ergs) (Derry et al. 2003; Vignali et
al. 2006; Ptak et al. 2006) or with AGN luminosities in the mid-infrared and
faint optical/near-infrared emission (Martinez-Sansigre et al. 2005; Houck et
al. 2005) are the two alternative approaches which have been proven to be efficient in discovering
large samples of QSO2s candidates. 
  
However, given the faintness in the 0.5-10
keV band  ({$F_{\rm 0.5-10}$ \simlt 10$^{-14}$ \cgs) of most of the QSOs candidates
detected so far, only for a handful of these objects has been possible to perform meaningful
X-ray spectroscopy and, therefore, obtain tight constraints on the absorbing
column density and, in general, on their X-ray properties, which are largely unexplored so far (e.g. Akiyama et al. 2002;  Severgnini et al. 2006; Piconcelli et al. 2007a).\\  

In this paper, we present a $\sim$40 ks \xmm~observation of \3c, a radio galaxy  with Fanaroff-Riley II edge-brightened morphology at
$z$ = 0.1848 (Riley \& Pooley 1975). \3c~was initially classified as a
broad line radio-galaxy owing to the presence of a weak broad
H$\alpha$ emission line in its optical spectrum (Grandi \& Osterbrock
1978).   However, Antonucci (1982) and Antonucci (1984) found that
both the optical/UV continuum  and  the broad H$\alpha$ emission are
highly polarized at a position angle perpendicular to the radio axis
(with evidence that the broad line is more polarized than the
continuum), while the narrow emission lines are very strong (L$_{\rm
[OIII]}$ $\sim$1.5 $\times$ 10$^{43}$ \ergs) and unpolarized.
Furthermore, indications of strong extinction can be inferred by the
absence  of the blue-UV emission bump and the large
broad-line Balmer decrement.  On the basis of these findings,
Antonucci \& Barvainis (1990) interpreted the polarization in terms of
scattered light from the obscured continuum and broad line region  by
a population of hot electrons and/or opaque dust clouds
(e.g. Kishimoto et al. 2001 for further details) and proposed a narrow
line radio galaxy (NLRG) classification for this source.  Tran et
al. (1995, T95 hereafter) presented the analysis of high
signal-to-noise (S/N) Keck spectropolarimetric data for \3c~and
confirmed  that the broad lines are caused by scattered light (Young
et al. 1998; Cohen  et al. 1999).  Applying corrections for reddening
and starlight contamination, T95 inferred an absolute magnitude of
M$_V$ $\leq$ $-$24.2, i.e. well within the quasar range.  All these
pieces of evidence indicate that \3c~harbors a buried luminous quasar
in the nucleus i.e. this source is a Type 2 quasar.  Although affected by a low
S/N ratio, the {\it ASCA} spectrum of \3c~(Sambruna et al. 1999)
provided further support to this classification suggesting the
presence of strong absorption (\nh~$\sim$ 10$^{23}$ \cm2), an
ultra-flat continuum ($\Gamma \sim$ 0.07) and a 2-10 keV luminosity of
$\sim$ 10$^{44}$ \ergs.\\

We adopt a $\Lambda$CDM cosmology with  $H_0$=70 km
 s$^{-1}$ Mpc$^{-1}$ and  $\Omega_\Lambda$ = 0.73 (Spergel et al. 2007). For \3c~at $z$ = 0.1848,
1 arcsecond corresponds to a physical scale of 3.1 kpc.

\section{\xmm~observation and data reduction}
\label{obs}

\3c~was observed by \xmm~(Jansen et al. 2001 and reference therein) on
 April 24, 2006 for $\sim$40 ks.
Data were reduced with SAS 6.5 using standard procedures.
X--ray events corresponding to patterns 0--12(0--4) for the
\mos(\pn)~cameras were selected. The event lists were filtered to
ignore periods of high  background flaring  according to the method
presented in Piconcelli et al. (2004) based on the cumulative
distribution function of background light curve count-rates. 
After this data cleaning, we obtained a net
exposure time
of 34.5 ks for the \pn~and of $\sim$38 ks for the two
\mos~cameras.
The source spectra were extracted 
using a circular region of 25 and 32  arcsecs radius for \pn~and
\mos, respectively, centered on the peak of the X-ray emission at
$\alpha_{2000}$ = 10$^{h}$01$^{m}$49.8$^{s}$ and
 $\delta_{2000}$ = $+$28$^{\circ}$47$^{\prime}$08$^{\prime\prime}$.
The \pn(\mos) background was estimated from a source--free region with a
 radius of 66(50) arcsec in the same chip.  
Response matrices and ancillary response
files were generated using the RMFGEN and ARFGEN tools in the SAS.
Combined \mos~spectrum and
response matrix were created. Events outside the 0.5-10 keV range were discarded
 in the \pn~spectrum, while we ignored \mos~data below the 0.8 keV due to the
 presence of cross-calibration uncertainties between the \mos~cameras (Kirsch 2006).
Both  \pn~and \mos~spectra were grouped to have a minimum of 20 counts per bin
to apply the  $\chi^{2}$ minimization technique, and fitted simultaneously.

During the observation the source flux remained steady, with
no variation exceeding 2$\sigma$ from the average count-rate level in both
soft- and hard-X-ray band and no significant spectral changes.
Therefore our spectral analysis was performed on the spectrum integrated
over the whole good exposure time.

\begin{figure}
\begin{center}
\includegraphics[width=5.5cm,height=7.5cm,angle=-90]{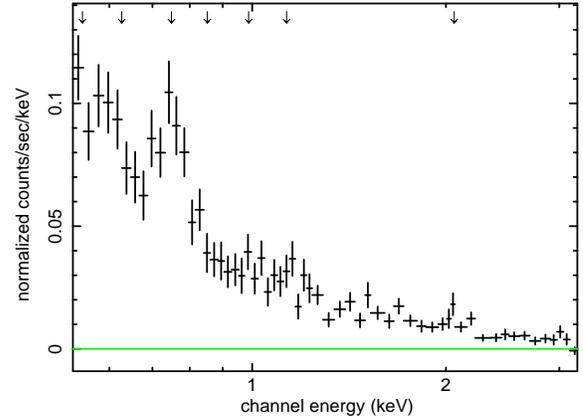}
\caption{The soft X-ray portion of the \pn~spectrum of \3c~when the hard X-ray {\it baseline} model 
  (i.e. absorbed power-law continuum $+$ narrow \feka~line) 
is applied to the 0.5--10 keV band. A large soft X-ray
excess dominated by several emission line-like features is seen.  The arrows  indicate the observer-frame energy of the significantly detected emission lines as listed in Table 2.}
\end{center}
\label{f:soft}
\end{figure}
\begin{table*}
\caption{Best-fit spectral parameters of the \epic~spectrum. 
See Sect.~\ref{spex} for details.}
\label{tab:fit}
\begin{center}
\begin{tabular}{cccccccccc}
\hline\hline\\
\multicolumn{1}{c} {Model}&
\multicolumn{1}{c} {$\Gamma$}&
\multicolumn{1}{c} {N$_{\rm H}$}&
\multicolumn{1}{c} {$f_s$}&
\multicolumn{1}{c} {$\Gamma_{\rm Soft}$}&
\multicolumn{1}{c} {$k$T}&
\multicolumn{1}{c} {Flux}&
\multicolumn{1}{c} {Luminosity}&
\multicolumn{1}{c} {EW$_{\rm K\alpha}$}&
\multicolumn{1}{c} {\xdof}\\
 (1)&(2)&(3)&(4)&(5)&(6)&(7)&(8)&(9)&(10)\\\hline\\ 
(A)&1.53$^{+0.03}_{-0.04}$&3.12$^{+0.17}_{-0.17}$&0.05&$\equiv \Gamma$&0.63$^{+0.06}_{-0.07}$&0.83/13.27&0.06$^\ast$/2.61&139$^{+42}_{-40}$&282(254)\\
&&&&&0.17$^{+0.01}_{-0.01}$&&&&\\
(B)&1.71$^{+0.05}_{-0.04}$&3.69$^{+0.19}_{-0.18}$&0.03&1.09$^{+0.13}_{-0.09}$&
$-$&0.85/13.18&1.62/2.97&135$^{+43}_{-41}$&255(246)\\
\hline
\end{tabular}\end{center}
The columns give the following information: (1) model (A: {\it baseline} model $+$ two {\tt mekal} components $+$ unabsorbed PL, B: {\it baseline} model $+$ soft X-ray Gaussian lines $+$ unabsorbed PL); 
(2) photon index of the absorbed continuum power law;
(3) the column density of the absorber (10$^{23}$ \cm2); 
(4) ratio between the normalization of the unabsorbed versus absorbed PL component at 1 keV; 
(5) photon index of the unabsorbed soft X-ray power law;
(6) the temperature (keV) of the thermal plasma component ({\tt mekal});
(7) the 0.5--2/2--10 keV observed flux (10$^{-13}$ \cgs); 
(8) the 0.5--2/2--10 keV luminosity  (10$^{44}$ \ergs); 
(9) EW of the \feka~line (eV); 
(10) $\chi^2$ and number of degrees of freedom.
$^\ast$ Luminosity of the thermal emission component in the 0.5--2.0 keV band.
\end{table*}
\subsection{Spectral analysis}
\label{spex}
All fits were performed using the
XSPEC package (v11.3) and included absorption due to the line-of-sight Galactic column
density of \nh~= 1.91 $\times$ 10$^{20}$ \cm2~(Dickey \& Lockman 1990). 
Hereafter, the quoted errors on the derived model
parameters correspond to a 90\% confidence level for one interesting
parameter (i.e. $\Delta\chi^2$ = 2.71). 
Best-fit parameter values are given in the source-frame, unless otherwise specified.

As a first step, we fitted the \epic~spectra at energies above 3 keV
using a simple power-law (PL) model  in order to  estimate the
``hardness'' of the X-ray continuum in \3c.  This fit was very  poor
(\xnn~$\sim$3), with a ultra-flat photon index of $\Gamma$ $\approx$ $-$0.4
denoting the complexity of the X-ray emission in this source and, in
particular,  the presence of strong absorption.  We then fitted the
3--10 keV spectrum  with a model consisting of a combination of a PL
(representing the primary continuum) absorbed by neutral matter at the
redshift of the source and a  narrow
Gaussian line to account for the \feka~emission at 6.4 keV.
This fit (hereafter indicated as {\it baseline} model) provides an excellent representation of the hard-band data, i.e. \xrd~= 1.04(162),
revealing an  X-ray primary continuum with slope $\Gamma$ $\approx$ 1.5 affected by a large amount of intrinsic absorption (\nh~$\approx$ 3$\times$ 10$^{23}$ \cm2).

\begin{figure*}
\begin{center}
\includegraphics[width=6cm,height=8cm,angle=-90]{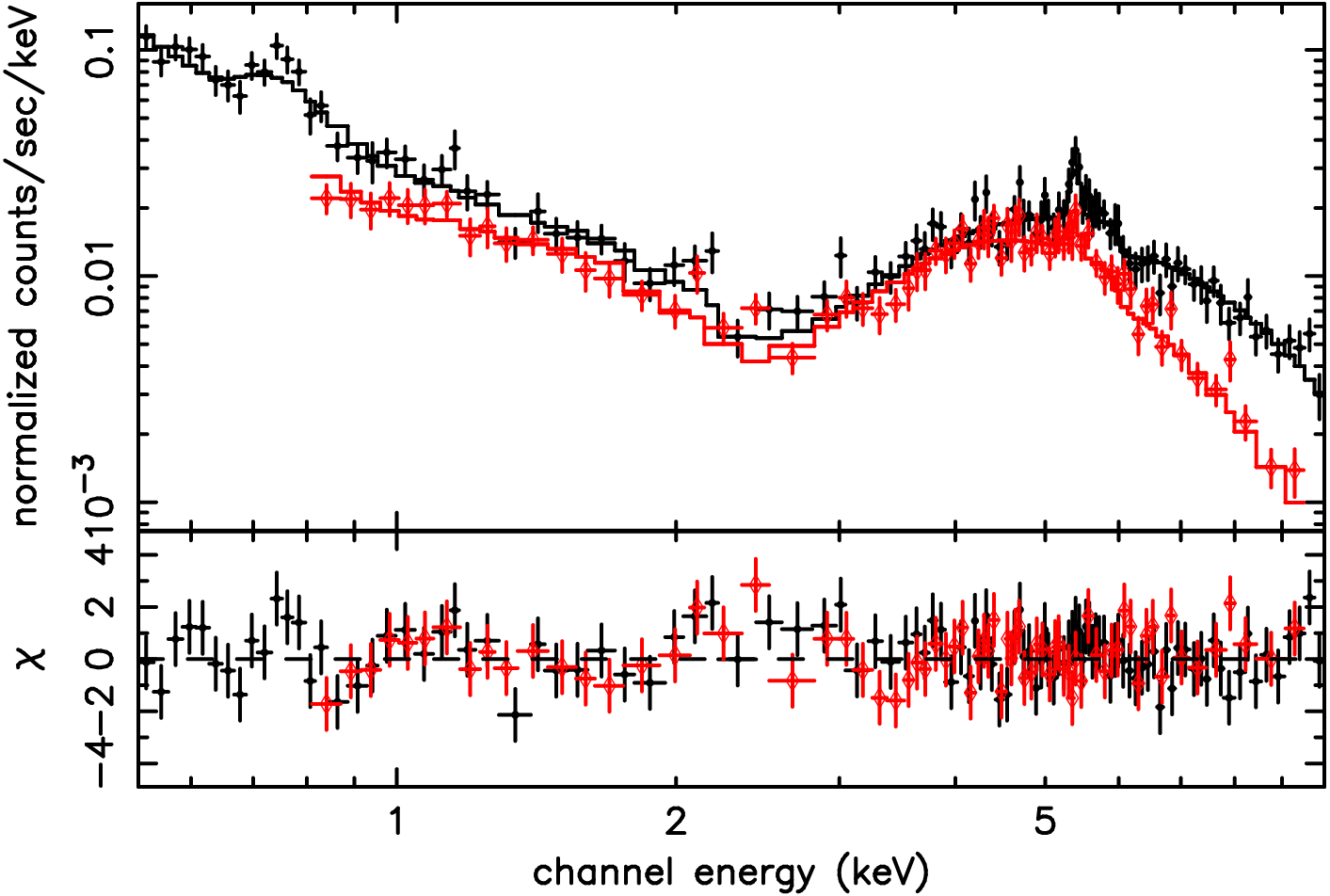}\hspace{0.5cm}\includegraphics[width=6cm,height=8cm,angle=-90]{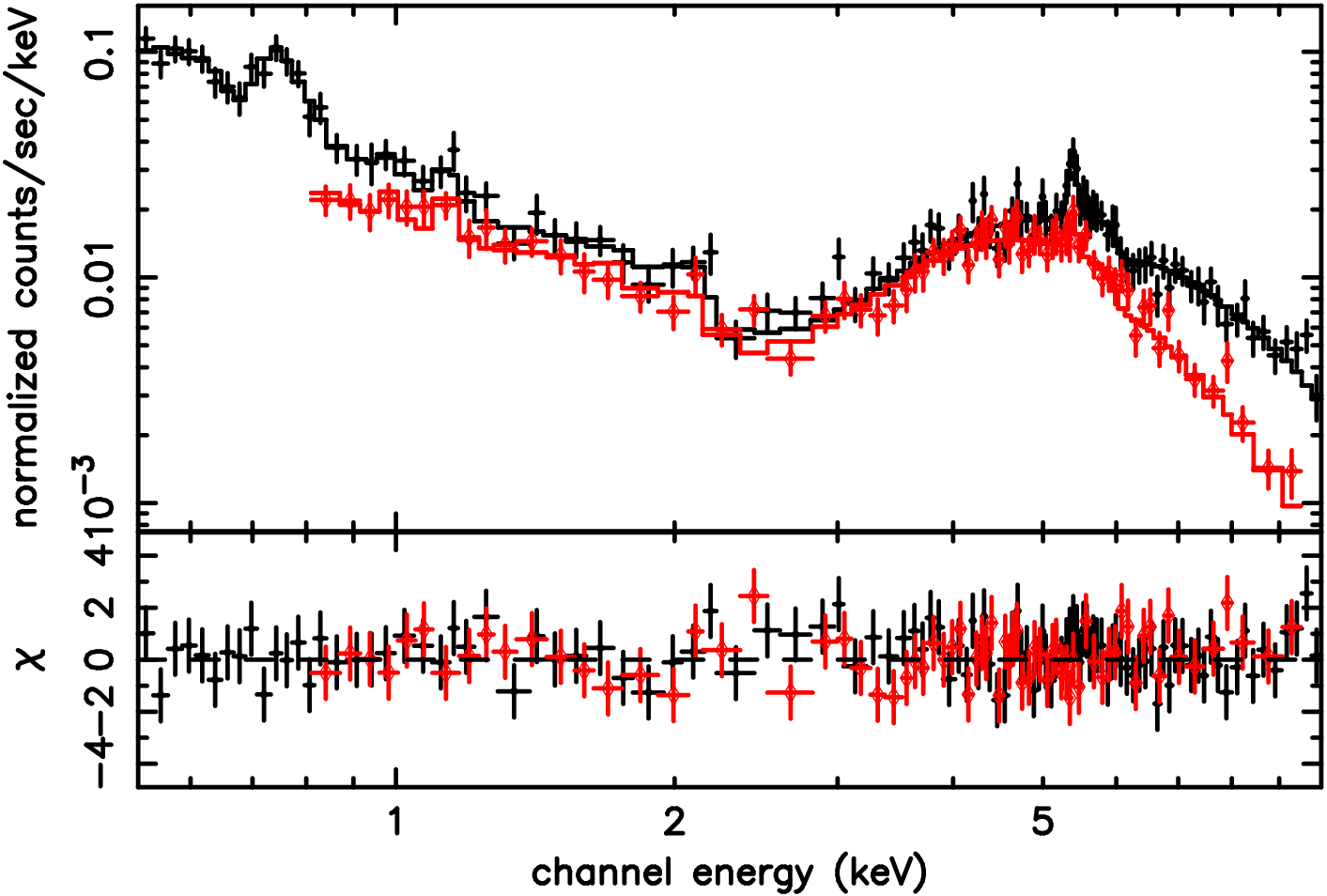}
\caption{{\bf(a)}--{\it Left:} \xmm~\pn~(top) and \mos~(bottom) spectra of
  \3c~when the model A is applied. 
The lower panel shows the deviations of the observed data from the model in
unit of standard deviation. {\bf(b)}--{\it Right:} The same as in the left
figure but for model B. See Table 1 for the best-fit parameters of each model.}
\end{center}
\end{figure*}

  The \feka~emission line  shows a
rest-frame energy of  E$_{\rm K\alpha}$=6.38$^{+0.03}_{-0.04}$ keV and
an equivalent width EW$_{\rm K\alpha}$ $\approx$ 140$\pm$40 eV. We
obtained an upper limit on the line width of $\sigma_{\rm K\alpha}$
$<$ 120 eV. 

The extrapolation of this spectral model to the soft X-ray band showed
the presence of  several narrow emission features along with a smooth
excess below $\sim$3 keV (e.g. Fig.~1).  Such narrow features in the
soft X-ray portion of the spectrum were initially modelled by the addition of a 
thermal plasma component ({\tt mekal} in XSPEC) to the hard X-ray {\it baseline}
model, under the hypothesis of an origin from starburst activity. This
fit was  completely unable to model the data at $E$ $<$ 3 keV with an associated
 \xnn~$\sim$6.  We then  added a second {\tt mekal} component
yielding a reasonable fit (\xrd~=1.16(255)).  However, some large
data/model residuals remained around 0.8-0.9 keV
(rest-frame). Furthermore, the temperature (and the 0.5-10 keV luminosity of 
$\approx$ 2 $\times$ 10$^{43}$ \ergs~as well) measured for one of the thermal components, 
i.e. kT$\sim$8 keV, seems
unlikely for a starburst region (Franceschini et al. 2003). It can be
an artifact of the fit related to the difficulty of fitting the soft
excess in the $\sim$1.5--3 keV band. Accordingly, we included an additional unabsorbed PL
component fixing its photon index to that of the absorbed PL, as
expected in case of scattered emission.
 The presence of the additional PL is required by the
data at the \simgt~99.8\% confidence level.   The temperatures of the two thermal components
were $k$T$\approx$ 0.2 and $k$T$\approx$0.6 keV, respectively.
The application of this model (e.g. Table 1, model A)  to
the data produces a  pretty acceptable fit to the \epic~data with an
associated \xrd~= 1.11(254) but leaves evident line-like positive
residuals below 1 keV  (see Fig.~\ref{f:soft}).
 

We then performed a phenomenological fit adding to the {\it baseline}
model seven narrow Gaussian lines to account for the emission features
observed in the $\sim$0.5--2 keV range.  Each  Gaussian line is required at a significance
level of P$_F$ $>$99.9\% (except for the line at 2.44 keV with P$_F$ =
98.5\%).  
 However, due to the potential line blending with adjacent emission lines such a detection significance should be considered with care and, in some cases, it may represent only an upper limit.
The photon index of the unabsorbed PL ($\Gamma_{\rm soft}$)
was left free to vary since blends of unresolved emission lines appear
as a pseudo-continuum when observed at the spectral resolution of
\epic. This PL  component also accounts for a
mixture of emissions caused by,  i.e., the electron-scattered fraction
of the primary continuum, thermal and/or photoionized plasma, the
relativisic jet and Compton reflection off ionized matter (see
Miniutti et al. 2007; Evans et al. 2006a; Turner et al. 1997), as
observed in most of X-ray obscured AGNs. 
We yielded a very good description of the spectrum of
\3c~(e.g. model B in Table 1), with an associated \xrd~= 1.04(246).
The confidence contours for $\Gamma$ versus \nh~are plotted in
Fig.~\ref{f:cont}.  The ratio between the normalizations of the
unabsorbed and  absorbed PL is $f_s$ $\approx$ 0.03.  In Table 2 
the best-fit parameters are listed  together with the likely identification
for each of the soft X-ray  lines. 

\begin{table}
\caption{Best-fit spectral parameters for the emission lines detected in the soft X-ray band.} 
\label{tab:fit}
\begin{center}
\begin{tabular}{cccc}
\hline\hline\\ \multicolumn{1}{c} {Energy}&\multicolumn{1}{c}
 {Intensity}&\multicolumn{1}{c} {P$_F$}&
 \multicolumn{1}{c} {Identification}\\
 (1)&(2)&(3)&(4)\\\hline\\ 
0.65$^{+0.01}_{-0.01}$&17.2$^{+4.9}_{-2.7}$&$>$99.9&OVIII Ly$\alpha$\\ 
0.74$^{+0.01}_{-0.02}$&11.1$^{+2.1}_{-1.9}$&$>$99.9& OVII RRC, Fe XVII 3s-2p\\
0.89$^{+0.01}_{-0.01}$&12.3$^{+1.2}_{-1.9}$&$>$99.9& OVIII RRC, NeIX He$\alpha$,\\
&&&Fe XVIII 3d-2p\\
1.02$^{+0.01}_{-0.02}$&1.4$^{+0.6}_{-0.9}$&$>$99.9& NeX Ly$\alpha$, Fe XXI 3d-2p\\
1.17$^\dag$&1.8$^{+0.6}_{-0.5}$&$>$99.9& FeXXIV L, NeX Ly$\beta$,\\
&&&  NeIX RRC\\
1.34$^\dag$&1.4$^{+0.5}_{-0.5}$&$>$99.9& MgXI He$\alpha$\\
2.44$^\dag$&0.4$^{+0.3}_{-0.1}$&98.5& SXV He$\alpha$\\
\hline
\label{t:lines}
\end{tabular}
\end{center}
The columns give the following information:
(1) Energy of the line (keV); (2)
  Intensity of the line (10$^{-6}$ photons cm$^{-2}$ s$^{-1}$); (3) Significance of the fit improvement after including the {\it Gaussian} line in the fitting model (based on
  $F$-test,  see Sect. 2.1 for further details on this significance value); (4) Likely identification (e.g. Brinkman et al. 2002; Guainazzi
  \& Bianchi 2007). $^\dag$ Indicates that the parameter has been fixed to the best-fit value
deduced by the fit.
\end{table}
Finally, the observed flux in the 2--10 keV band is \fhx~$\sim$1.3
$\times$ 10$^{-12}$ \cgs, which is consistent with the value measured
with {\it ASCA} in 1994 (Sambruna et al. 1999).

\section{Discussion}
\label{discussion}

\begin{figure}
\begin{center}
\includegraphics[width=6cm,height=8cm,angle=-90]{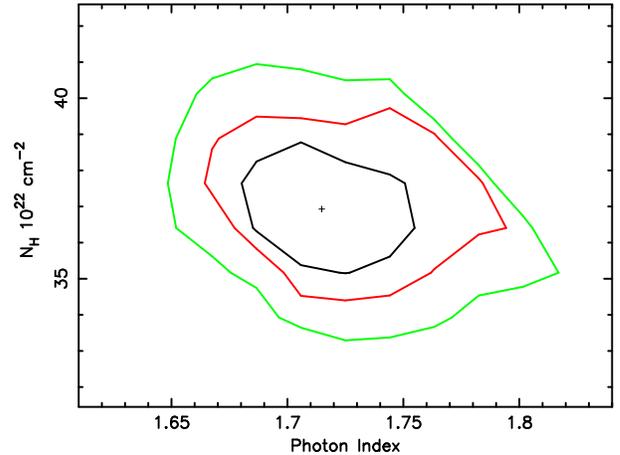}
\caption{Confidence contour plot showing the intrinsic
  rest--frame column density against the hard X-ray continuum photon index.  The contours are at 68\%, 90\% and 99\% confidence levels for two interesting parameters.}
\end{center}
\label{f:cont}
\end{figure}
\subsection{The Hard X--ray spectrum and the Fe K$\alpha$ emission line}
\label{s:hard}

The \epic~observation presented here is the first good-quality X-ray
spectrum of \3c~and, in general, it represents one of the best X-ray
observation collected for a genuine QSO2 so far.  
Our analysis reveals the presence of a complex spectrum, largely dominated by strong
obscuration. The X-ray continuum shows a
slope of $\Gamma$ $\sim$ 1.7 and is absorbed by Compton-thin
(\nh~$\sim$ 3 $\times$ 10$^{23}$ \cm2) neutral gas.  Since the
intrinsic 2--10 keV luminosity is \lum~$\sim$ 3 $\times$ 10$^{44}$
\ergs, \3c~satisfies all the conditions to be considered a genuine
Type 2 quasar.

The value of the photon index of the continuum is consistent with the
 average value found for other lobe-dominated\footnote{Lobe-dominated
 radio-loud sources are those with a rest frame ratio (the so-called
 {\it core dominance} R$_c$) of radio core to lobe flux density of
 log R$_c$ $<$0 (Orr \& Brown 1982). According to Sambruna et
 al. (1999) \3c~has a log R$_c$ = $-$1.34.} FRII quasars,
 i.e. $\langle\Gamma\rangle$ $\sim$ 1.6--1.7 (Belsole et al. 2006;
 Piconcelli et al. 2005; Grandi et al. 2006).  Harder photon indices
 of $\Gamma$ $\sim$ 1.4--1.5 are indeed usually observed in the
 core-dominated, more-aligned radio-loud AGNs, the so-called {\it
 Blazars},  whose featureless X-ray spectrum is dominated by
 relavistically-beamed inverse-Compton jet emission (e.g. Fossati et
 al. 1998; Donato et al. 2005).  The \xmm~measurement implies that the
 jet contribution to the broad-band X-ray emission must be marginal as
 suggested by the detection of many soft X-ray emission lines  as well
 as a prominent \feka~line at 6.4 keV in its spectrum. 
 It is worth noting that this is the first time that such emission line has been detected at a statistically significant level in \3c.
 The best-fit
 rest-frame energy of the \feka~line is  E$_{\rm
 K\alpha}$=6.38$^{+0.03}_{-0.04}$ keV, corresponding to low ionization
 states, i.e. FeI--XVI (Kallman et al. 2004). The EW$\approx$140 eV of
 this line matches well with the value predicted  by theoretical
 calculations (e.g. Awaki et al. 1991; Leahy \& Creighton 1993;
 Ghisellini et al. 1994) for a \feka~line produced by transmission
 through an absorbing screen with a column of \nh~$\sim$3--4  $\times$
 10$^{23}$ \cm2, as that measured along the line of sight to the
 nucleus of \3c~by the \xmm~observation.  This suggests that such
 Compton-thin absorber is the responsible for both obscuration and
 fluorescent narrow \feka~emission.  However, the value of the EW is
 also consistent with reflection off distant Compton-thick matter
 (Guainazzi, Matt \& Perola 2005). The cold reflector could be then located at the far
 inner side of the molecular {\it torus} invoked in the AGN  unified
 models (Antonucci 1993; see also Elitzur \& Shlosman 2006) and  seen along an unobscured line of sight.  This
 scenario implies the simultaneous presence of two circumnuclear
 obscuring gases, one Compton-thick (the reflector) and one
 Compton-thin (the absorber), as recently discovered in some Seyfert 2
 galaxies showing large X-ray spectral variations (Matt et al. 2003;
 Risaliti et al. 2005; Piconcelli et al. 2007b). If this is the case,
 the Compton-thin matter should cover only a small fraction of the sky
 (as seen by the X-ray primary source) in order not to produce too
 much iron K$\alpha$ emission (Matt 2002). 
  Furthermore the \feka~line should be
 accompanied by a Compton-reflection continuum emission peaking at
 $\sim$30 keV (Matt, Perola \& Piro 1991; Krolik et al. 1994).  Unfortunately, the limited
 \epic~bandpass does not allow to constrain meaningfully the strength
 of this spectral component in \3c, and, in turn, test the validity of
 this hypothesis. Future spectroscopy in the 10--60 keV range, say with {\it Simbol-X} (Ferrando et al. 2006), will be therefore crucial 
to disentangle between a transmission or a reflection origin for the
\feka~line in this QSO2.

Young et al. (1998) inferred a dust extinction toward \3c~of $A_V$ = 60
mag, which corresponds to an absorbing column density of $\approx$ 1.1
$\times$ 10$^{23}$ \cm2, if the standard Galactic gas-to-dust ratio is
assumed (i.e.  \nh/$A_V$=1.79$\times$ 10$^{21}$ \cm2~mag$^{-1}$,
Predehl \& Schmitt 1995),  which is a factor of $\sim$ 3.4 lower than
observed.  This mismatch has been observed in many X-ray sources as reported by Maiolino et al. (2001), that  proposed
the existence of dust grains larger than in the diffuse Galactic
medium which do not redden efficiently the optical emission.
Alternatively,   Weingartner \& Murray (2002) suggested that X-ray
absorption and optical extinction  occur in distinct regions, with the
X-ray absorber located inside the dust sublimation radius.

\subsubsection{The [Mg VIII] to 2--10 keV luminosity ratio}
Imanishi (2006) reported the detection of a strong high-excitation
emission line at 3.028 $\mu$m due to [Mg VIII], a hallmark of the
presence of a powerful AGN in this source.  We derive a [Mg VIII] to
2--10 keV luminosity ratio \mgx~=0.004 assuming a  [Mg VIII]
luminosity $L_{Mg^{7+}}$ = 1.2 $\times$ 10$^{42}$ \ergs.  Noteworthy,
such a  ratio is similar to that inferred for the archetypical
Seyfert 2 galaxy Circinus, i.e. \mgx~=0.005 using a $L_{ Mg^{7+}}$ =
4.75 $\times$ 10$^{39}$ \ergs (Sturm et al. 2002) and a \lum~=
10$^{42}$ \ergs~(Matt et al. 1999), despite the fact that the
$L_{Mg^{7+}}$ of the QSO2 is approximatively a factor of 250 larger
than in Circinus.  A value of \mgx~=0.004 is also found for the
heavily Compton-thick source NGC 1068 ($L_{Mg^{7+}}$ = 3.1 $\times$
10$^{40}$ \ergs;  Marco \& Brooks 2003) if we adopt  a \lum~=  7.7
$\times$ 10$^{42}$ \ergs, i.e. the value determined by Levenson et
al. (2006) from the \feka~line luminosity (see also Panessa et
al. 2006).  The possibility of using $L_{Mg^{7+}}$ as a proxy of the
intrinsic X-ray luminosity of an obscured active nucleus deserves
further investigations, therefore a larger sample of AGN is needed to
confirm this relationship on a more sound statistical
ground. Unfortunately, an accurate measurement of the intensity [Mg
VIII] emission  line at  3.028 $\mu$m is available to date in
literature only for a handful of sources.

\subsection{The Soft X--ray emission}
\label{s:soft}
The emission lines detected for the first time at \simlt~2 keV in the
\xmm~spectrum of \3c~(see Table 2) rule out the hypothesis that in
this  radiogalaxy the bulk of the soft X-ray emission arises from
jet-related non-thermal emission, as suggested by Evans et al. (2006a)
and Belsole et al. (2006) by the analysis of X-ray
observations of narrow-line radio-loud AGNs.  It is worth noting that
a simple fit with an unabsorbed PL in the 0.5--2 keV band yields a very steep
photon index ($\Gamma \sim$ 2.6), which is in agreement with slopes of
the soft X-ray PLs in absorbed FR II radio galaxies reported by
Belsole et al. (2006).

Furthermore, as mentioned in Sect.~\ref{spex}, the 0.5--2 keV
luminosity of $\approx$ 5.8 $\times$ 10$^{42}$ \ergs~of the
collisionally-heated plasma component ({\tt mekal} in XSPEC) inferred
by model A is too large to be associated with starburst regions, which
typically show 10$^{39}$ \simlt~L$_{0.5-2}$ \simlt~a few times
10$^{41}$ \ergs~(e.g. Ptak et al. 1999; Franceschini et al. 2003).
Using the relationship between  far infrared (FIR) and soft X-ray
luminosity reported by Ranalli et al. (2003) for a large sample of
star-forming galaxies, we estimate a starburst luminosity of
L$^{SB}_{0.5-2}$ \simlt~3.3~$\times$ 10$^{41}$ \ergs, i.e. a contribution
\simlt~4\% to the total soft X-ray observed luminosity in \3c~(L$^{Obs}_{0.5-2}$ = 8.2
$\times$ 10$^{42}$ \ergs),  assuming a FIR luminosity of L$_{\rm FIR}$
\simlt~1.54 $\times$ 10$^{45}$ \ergs. 
 According to Kennicutt (1998) the  FIR luminosity of \3c~implies a star formation rate 
of SFR $<$ 70 M$_\odot$ yr$^{-1}$, while a L$_{0.5-2}$ = 5.8 $\times$ 10$^{42}$ \ergs~translates 
into a SFR $\sim$ 1300 M$_\odot$ yr$^{-1}$ (Ranalli et al. 2003), i.e. a very large value which is 
observed only in the most massive nuclear starbursts  triggered by strong tidal interactions and 
mergers of galaxies. 
 This finding is in agreement with the results of Imanishi (2006) that found
in the 3--4 $\mu$ ($L$-band) spectrum of \3c~3.3 $\mu$m emission PAH
features  more than an order of magnitude weaker than those of
typical starburst galaxies.

As the soft X-ray emission is not affected by absorption, the location
of the emitting gas in this QSO2 must be far away from the obscuring
gas  intercepting the primary continuum.

The soft X-ray emission lines found by \xmm~can be associated (bearing
in mind the EPIC resolution and the error bars of the energy values
listed in Table 2) with hydrogen- and helium-like lines of the most
abundant light metals, from oxygen to sulfur. This result is very
similar to those reported from the \xmm~and \chandra~observations of
obscured radio-quiet Seyfert-like AGNs (e.g. Guainazzi \& Bianchi 2007
and references therein), in which the soft X-ray emission is dominated
by a wealth of emission lines and spatially coincident with the
extended [OIII] emission (Iwasawa et al. 2003; Bianchi et al. 2006). High-resolution X-ray grating spectroscopy of this soft
X-ray emission provides unambiguous evidence that most of it is
produced in a photoionized outflowing gas (e.g. Kinkhabwala et
al. 2002; Guainazzi \& Bianchi 2007).  On the other hand, the exact physical mechanism generating
the emission-line spectra is still debated.  Both  AGN-photoionization and ``local''
photoionization due to gas heated by high-speed shocks driven by
nuclear outflows (i.e. the radio jet) in the interstellar medium (Fu
\& Stockton 2006; Evans et al. 2006b)  are viable
mechanisms.

This scenario could likely hold also in \3c~even if this source is a
radio-loud QSO2. In particular, the detection of two prominent lines
at 0.74 and 0.88 keV bolsters this hypothesis since they may be
associated with narrow OVII and OVIII Radiative Recombination
Continuum (RRC), respectively, which are hallmarks of recombination in a
low-temperature photoionized plasma (Liedahl 1999).   However, better
quality high-resolution spectral data are necessary before a firm
conclusion can be drawn on the exact origin of these emission lines.
Due to the weakness of \3c~in the soft X-ray band (\fsx~$\approx$8
$\times$ 10$^{-14}$ \cgs) no meaningful RGS data were collected during
the present \xmm~observation. This does not allow to  estimate the
possible contribution arising from collisionally-ionized plasma.

The discovery of a rich emission-line soft X-ray spectrum in \3c~is
noteworthy since this is the first time it has been found in a Type 2
and, even more remarkable, radio-loud QSO.  In fact, only very
recently   Sambruna et al. (2007) and Grandi et al. (2007) reported
the first clearcut example of a radio-loud AGN with a soft X-ray
spectrum dominated by emission lines on the basis of the
\xmm~observation of the heavily absorbed FRII radio galaxy 3C 445
(e.g. also Young et al. 2002).  These findings suggest that inner
circumnuclear regions in radio-loud (in particular those with high
excitation optical emission lines, which usually show a FRII radio
morphology, e.g. Grandi \& Palumbo 2007; Ballantyne 2007) and
radio-quiet AGNs could have similar geometrical and physical
properties.\\

Finally, according to model B the unabsorbed PL has a photon index of
$\Gamma_{\rm Soft}$ = 1.09$^{+0.13}_{-0.09}$. 
 The
0.5--2 keV luminosity of this spectral component is  $L_{0.5-2}$
$\approx$ 4  $\times$ 10$^{42}$ \ergs, which represents about 2--3\% of
the intrinsic luminosity of \3c~in the soft X-ray band ($\approx$ 1.6  $\times$ 10$^{44}$ \ergs).
 This emission could be
due to a blend of nuclear scattered emission arising from the
line-emitting warm photoionized plasma (and/or the electron mirror
located interior to the NLR proposed by T95 on the basis
of spectropolarimetric data) and unresolved X-ray emitting knots of
the kpc-scale jet (which usually show flat X-ray slope,  $\Gamma$ $\approx$ 1.0--1.3; e.g. Sambruna et al. 2004), plus a possible, marginal 
(i.e. $<$ 8\%, assuming a starburst luminosity of
L$^{SB}_{0.5-2}$ $<$ 3.3~$\times$ 10$^{41}$ \ergs~estimated from the FIR luminosity of \3c; e.g. Ranalli et al. 2003) contribution from X-ray binaries.
 

\section{Summary}

In this paper, we present the analysis of the \xmm~observation of
\3c~collected in 2006. These data provide the most detailed
description to date of the X-ray spectral properties of this FRII
radio galaxy, which has been long known for hosting a hidden quasar
at its center (Antonucci \& Barvainis 1990).

The quality of the data allows to shed light on many controversial
issues about the characteristics of the X--ray emission in \3c, i.e. the exact shape of
the continuum, the iron K fluorescent emission and the nature of the
soft X-ray spectral component, left unsettled by the low signal-to-noise ratio
1994 {\it ASCA} observation.  We measure a \lum~$\approx$3 $\times$
10$^{44}$ \ergs, whereby \3c~falls well within the  quasar X-ray luminosity
range (Fabian 2003).  At face value, the 0.5--10 keV spectrum of this
radio-loud QSO2 is typical  of a nearby Compton-thin Seyfert 2 galaxy. It is
dominated by heavy obscuration (\nh~$\sim$3.7  $\times$
 10$^{23}$ \cm2), with a strong, narrow
\feka~emission line at 6.4 keV and many emission lines from highly
ionized metals in the soft X-ray band.  Such a line-rich spectrum and
the steepness of the hard X-ray continuum, $\Gamma$ $\approx$ 1.7, reveal that the
non-thermal emission arising from the relativistic jet provides a
marginal contribution to the X-ray emission.  
The strength of the
\feka~line is fairly consistent with that expected in transmission from the observed obscuring screen 
with \nh~$\sim$4  $\times$
 10$^{23}$ \cm2. However, a line origin from reflection occuring at the AGN-illuminated, far inner surface 
of the Compton-thick {\it torus} cannot be rule out. This in turn would imply the presence of two
absorbing regions coexisting in this QSO2.

The detection of lines at 0.74 and 0.87 keV (i.e. the energy of the
narrow OVII and OVIII RRC, respectively) strongly indicates that most
of the soft X-ray emission originates from a photoionized plasma.  The
presence of this warm gas is particularly important since it suggests
that the circumnuclear environment in radio-loud and radio-quiet AGNs
share the same geometrical and physical properties. This remarkable
finding leads support to recent \xmm~results (e.g. Grandi et al. 2007;
Sambruna et al. 2007), which reported a similar scenario for 3C 445, another
X-ray luminous, obscured FRII radio galaxy.

\begin{acknowledgements}
We thank the anonymous referee for a number of helpful comments which improved the paper.
The authors  kindly thank  Roberto Maiolino  (OAR) for very useful discussions.
EP  is supported by an INAF Post-doctoral
Fellowship.
Based on observations obtained with XMM-Newton, an ESA science mission
with instruments and contributions directly funded by ESA Member
States and NASA. This research has made use of the NASA/IPAC Extragalactic Database (NED) which is operated by the Jet Propulsion Laboratory, California Institute of Technology, under contract with the National Aeronautics and Space Administration.
EP acknowledges financial contribution from contract ASI-INAF I/023/05/0.

\end{acknowledgements}

\end{document}